\documentclass[sigconf]{acmart}
\AtBeginDocument{%
  }

\usepackage{amsmath}
\usepackage{adjustbox}
\usepackage{float}

\copyrightyear{2025}
\acmYear{2025}
\setcopyright{acmlicensed}\acmConference[WSDM '25]{Proceedings of the Eighteenth ACM International Conference on Web Search and Data Mining}{March 10--14, 2025}{Hannover, Germany}
\acmBooktitle{Proceedings of the Eighteenth ACM International Conference on Web Search and Data Mining (WSDM '25), March 10--14, 2025, Hannover, Germany}
\acmDOI{10.1145/3701551.3703514} \acmISBN{979-8-4007-1329-3/25/03}

\begin{document}

\title{Writing Style Matters: An Examination of Bias and Fairness in Information Retrieval Systems }

\author{Hongliu CAO}
\email{hongliu.cao@amadeus.com}
\email{caohongliu@gmail.com}
\orcid{0000-0002-1326-8159} 
\affiliation{%
  \institution{Amadeus}
  \city{Nice}
  \country{France}
}


\begin{abstract}
The rapid advancement of Language Model technologies has opened new opportunities, but also introduced new challenges related to bias and fairness. This paper explores the uncharted territory of potential biases in state-of-the-art  universal text embedding models towards specific document and query writing styles within Information Retrieval (IR) systems.   Our investigation reveals that different embedding models exhibit different preferences for document writing style, while more informal and emotive styles are less favored by most embedding models. In terms of query writing styles, many embedding models tend to match the query style with the retrieved document style, but some show a consistent preference for specific styles. Text embedding models fine-tuned on synthetic data generated by LLMs display a consistent preference for certain style of generated data. These biases in text embedding based IR systems can inadvertently silence or marginalize certain communication styles, thereby posing a significant threat to fairness in information retrieval.  Finally, we also compare the answer styles of Retrieval Augmented Generation (RAG) systems based on different LLMs and find out that most text embedding models are biased towards LLM's answer styles when used as evaluation metrics for answer correctness.  This study sheds light on the critical issue of writing style based bias in IR systems, offering valuable insights for the development of more fair and robust models.
\end{abstract}




\maketitle

\section{Introduction}

Information Retrieval (IR) systems are pivotal in managing the current informational overflow, empowering users to obtain results more efficiently \cite{dai2024unifying}. The relevance of IR in everyday life is profound due to its use cases in diverse applications such as online web browsing, question-answering systems, digital assistants, etc. The central goal of IR systems is to identify and fetch data relevant to the user's query (often based on their embedding similarity scores)
\cite{hambarde2023information}. Text embeddings for information retrieval have undergone substantial transformations throughout the past decades, from the early phase of count-based sparse embeddings such as Term Frequency-Inverse Document Frequency (TF-IDF) \cite{manning2008introduction}, transitioning into the second phase characterized by static dense word embeddings, including Word2Vec \cite{mikolov2013efficient} and FastText \cite{bojanowski2017enriching}. The field then evolved into the third phase featuring contextualized embeddings like Bidirectional Encoder Representations from Transformers (BERT) \cite{devlin2018bert} and Embeddings From Language Models (ELMo) \cite{elmoneumann2018deep}, leading to the latest developments in the fourth phase encompassing universal text embeddings, such as General-purpose Text Embedding model (GTE) \cite{gteli2023towards}, Beijing Academy of Artificial Intelligence General Embedding (BGE) \cite{bgexiao2023c}, Large Language Model (LLM) based  universal text embedding \cite{e5mistralwang2023improving}, among others \cite{cao2024recent}.

With the widespread applications and use cases of ChatGPT, the remarkable capacities of LLMs have been highlighted, especially  in terms of conversational capabilities, instruction-following and in-context learning \cite{cao2024recent}, which has fundamentally revolutionized Information Retrieval. However, several critical issues of LLMs have also been identified, including lack of some domain knowledge and updated knowledge, and hallucinations. One solution to mitigate these limitations is Retrieval Augmented Generation (RAG) which uses text embedding to provide relevant contexts for LLMs to generate answers beyond their initial training data \cite{huang2024survey}. LLMs are also used to generate synthetic data to fine-tune text embedding models for IR systems \cite{e5mistralwang2023improving, lee2024gecko}.  
The fast adoption of LLMs in IR has also brought new challenges related to bias and fairness. These issues can undermine the reliability and robustness of IR systems and may inadvertently contribute to broader societal problems \cite{dai2024unifying}. 

Recently, considerable attention has been devoted to the bias and fairness related to machine learning models \cite{dai2024unifying, cao2023inclusive}. However, only a limited number of research initiatives have been concentrated on the potential bias present in text embedding models utilized for IR systems. 
Most existing studies ignore the fact that both human written documents and LLM-generated synthetic documents can have diverse communication or writing styles.
These variations in writing styles can introduce bias and unfairness into information retrieval systems in several ways. 
Firstly, if the text embedding models have been trained mainly on data in a particular writing style, they may be biased and struggle to accurately interpret and retrieve information from documents written in a different style. 
Secondly, the system may exhibit unfairness across different writing styles. For instance, if the system has been trained to associate higher relevance with documents that use complex sentence structures, it may favor such documents in the search results even if they are not the most relevant to the user's query while discriminating against documents in other writing styles. 
Finally, the writing style of queries could also introduce bias. If the system is trained with  queries based on specific writing styles, it may fail to answer queries written in other styles.

In this study, we fill the gap of the literature by examining the influence of varying writing styles on state of the art text embedding models with the following key contributions:
\begin{itemize}
    \item This work pioneers the study of bias and fairness in IR systems related to document and query writing styles.
    \item A comprehensive analysis and comparison is provided over diverse state of the art text embedding models with different fine-tuning data, fine-tuning strategies, objective functions, model size, backbones, etc.
    \item A novel method of analyzing and quantifying fairness in text embedding models is proposed, which reveals different writing style biases in different embedding models, as well as the impact of query styles on model fairness.
    \item The potential bias of  text embedding models towards  different LLM's answer styles (when used as evaluation metrics for answer correctness)  is also identified and analyzed.
\end{itemize}

\section{Related Works}
\textbf{Bias in Information Retrieval systems:} The issue of bias and unfairness in the field of Artificial Intelligence (AI) has been a subject of extensive research in past years. The recent incorporation of LLMs into information retrieval systems has augmented the complexity of identifying bias. This is primarily because LLMs are used in multiple stages in IR systems including synthetic data generation to augment the training data of IR systems, synthetic web content generation (input of IR systems),  LLMs as backbone of text embedding models, LLMs for re-ranking, LLMs as IR models and LLMs as judge to evaluate the IR system performance \cite{dai2024unifying, lee2024gecko, adlakha2023evaluating}. 

Many state of the art studies focus on the identification of bias on LLM-based IR systems such as RAGs. For instance, several research studies have underscored the fact that the sequence in which candidate documents or items are presented to LLMs has a substantial influence on the performance of LLM-based IR systems \cite{tang2023foundpos, hou2024largepos}. In \cite{wu2024faithful, tan2024blinded}, the tug-of-war between retrieved content and LLMs’ internal prior was identified: if the retrieved information significantly diverges from the LLM's internal knowledge, there's a decrease in its likelihood to favor the retrieved information while generating its response. When LLMs are used as judges for IR systems, numerous biases have been detected. Notably, these LLM judges have a tendency to neglect factual inaccuracies in an argument \cite{chen2024humansjudge} and demonstrate a preference for responses that they have generated or those originating from LLMs within the same category \cite{liu2023ego, liu2023llmsego}.

In both classic and LLM-based IR systems, text embeddings play an important role \cite{cao2024recent}, while only a few studies have focused on the potential bias and fairness in text embedding models for information retrieval. 
Recent research conducted by \cite{chen2024spiral} explored the immediate and long-term impacts of texts generated by LLMs on the performance of RAG systems. The findings suggest that the LLM-created content consistently excels in search rankings over human-written content, which consequently reduces the visibility and influence of human-generated  online contributions. In \cite{dai2023llmsinfoaccess}, the authors use LLMs to generate rewritten text copies of human written texts while preserving the same semantic meaning. They subsequently noted that neural text embedding models are biased towards LLM-generated texts.

\textbf{Universal text embeddings:} 
The rise in the popularity of text embedding techniques is evident across both industrial and academic sectors, a trend driven by their critical role in a variety of natural language processing tasks including information retrieval, question answering, semantic textual similarity, item recommendation, etc \cite{cao2024recent}. 
It has been long-standing goal in the field of text embeddings to develop  a universal model which is able to address a multitude of input text length, downstream tasks, domains and languages \cite{gteli2023towards}. With the increasing number and improved quality of diverse text datasets across different tasks, synthetic data generation, LLMs as backbones for embedding model as well as benchmarks with the focus on novel tasks and domain generalization such as the Massive Text Embedding Benchmark (MTEB) \cite{muennighoff2022mteb}, significant advancements have been achieved in the domain of universal text embeddings recently \cite{cao2024recent}.

The state of the art universal text embeddings can be divided into two main categories based on the backbone models they use: BERT-based and LLM-based models. BERT-based universal text embeddings use BERT or BERT-like contextual embedding models (e.g. xlm-roberta \cite{conneau2019unsupervised}) as backbones with the number of parameters typically ranging from 33M to 560M. 
GTE \cite{gteli2023towards} is one of the top performing  BERT-based universal text embeddings, which  used 800M text pairs for pre-training  and employed a varied blend of datasets from numerous sources for the supervised fine-tuning phase \cite{gteli2023towards}. 
On the other side, BGE \cite{bgexiao2023c} aimed to enhance the quality of training data by eliminating irrelevant text pairs and incorporating high-quality, multi-task labeled data for task-specific fine-tuning. 
Multilingual-e5-large-instruct \cite{e5instructwang2024multilingual}  utilized 1 billion assorted multilingual text pairs for contrastive pre-training. An additional 500K synthetic data generated by GPT-3.5/4 including 150K distinctive instructions across 93 languages combined with real-world data were used for fine-tuning \cite{e5instructwang2024multilingual}. mxbai-embed-large-v1 \cite{mrlkusupati2022matryoshka}  introduced a novel loss for learning data representations through a nested structure, thereby fostering flexibility in the learned representation.

A significant benefit of employing LLMs as backbones for text embedding is their extensive pre-training on large-scale web data, thereby eliminating the necessity for the contrastive pre-training phase \cite{cao2024recent}.
SFR-Embedding-2\_R \cite{SFR-embedding-2} fine-tunes Mistral7B \cite{jiang2023mistral} model with task-homogeneous batching, multi-task datasets and improved hard negatives.  
LLM2Vec-Llama-2-7b \cite{llm2vec} transforms decoder-only LLM (e.g. Llama-2-7b \cite{touvron2023llama}) to  a strong text embedding by enabling bidirectional attention and fine-tuning with Masked Next Token Prediction (MNTP) plus unsupervised contrastive learning. 
Based on Qwen2-7B \cite{yang2024qwen2}, gte-Qwen2-7B-instruct \cite{gteli2023towards} is fine-tuned with the use of bidirectional attention mechanisms as well as a large multilingual text corpus that covers various domains and scenarios.

\textbf{Summary:} 
In the recent literature, considerable attention has been devoted to the bias and fairness related to LLM applications while only a limited number of research initiatives have been concentrated on the potential bias present in text embedding models utilized for IR systems and QA systems. There are two main limitations of the existing bias and fairness studies: Firstly, the potential diversity in writing styles inherent in both human-written and LLM generated documents is overlooked. 
 Secondly, the text embedding models studied in existing literature are mostly classic ones, ignoring the recent advancement in universal text embeddings. 

\section{The bias and fairness related to document writing styles}
Individuals possess unique writing styles that could be influenced by factors such as their educational background, cultural influences, personal experiences, or linguistic preferences. These styles can vary significantly in terms of vocabulary usage, sentence structure, tone, etc. For example, some people may prefer short, concise sentences, while others may favor more complex and lengthy sentence structures. 
In order to study whether the state of the art text embedding models are biased towards certain document writing styles and discriminate against other writing styles, extensive experiments are designed in this section.

\subsection{Data}

\textbf{Writing styles:} 9 diverse writing styles are selected and compared in this work, including:
\begin{itemize}
    \item 
    Style-0: The default style of LLM (with the prompt 'Please rewrite the following text' from \cite{dai2023llmsinfoaccess}.)
\item  Style-1: Your writing style is formal, efficient, and concise, using professional language and focusing on facts, figures, and data.
\item Style-2: Your writing style is clear and using simple language, often avoiding idioms or complex sentences.
\item Style-3: Your writing style is informal, often includes emojis, abbreviations, and internet slang.
\item Style-4: Your writing style is polite, respectful, and somewhat formal. You use more traditional language and avoid using slang or abbreviations.
\item Style-5: Your writing style is formal, detailed, and precise manner with structured texts. You use technical language and focus on evidence-based arguments.
\item Style-6: Your writing style is energetic, motivational, and positive manner.
\item Style-7: Your writing style is friendly, casual, and empathetic manner with personal anecdotes
\item Style-8: Your writing style is expressive and emotive (passionate, engaging, empathetic). You use metaphors, analogies, and storytelling to convey your points.
\end{itemize}

\begin{figure}[h]
  \centering
  \includegraphics[width=\linewidth]{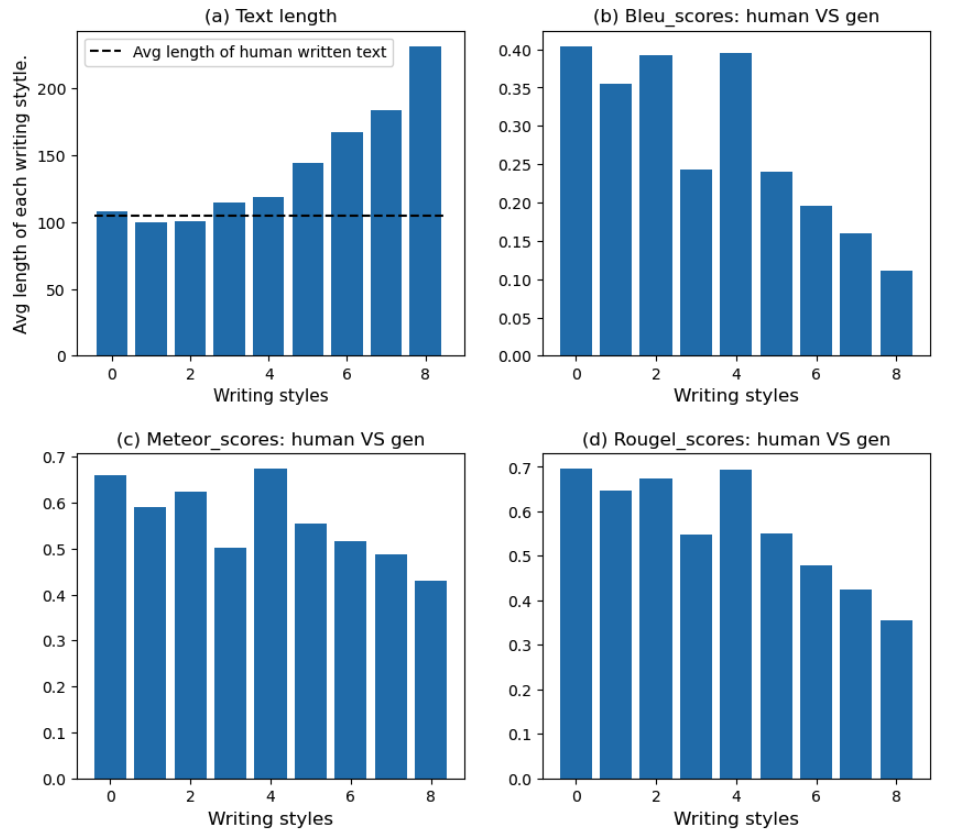}
  \caption{Basic statistics comparison between human written documents and generated documents in 9 different styles which corresponds to values 0-8 on X-axis: (a) comparing the text length of different styles with the average length of the original text (dashed black line); (b),  (c)  and (d) are  Bleu scores, Meteor scores and RougeL scores  between original text and 9 writing styles. } 
  \label{fig:stats}
\end{figure}

\textbf{Dataset:} Following the work of \cite{adlakha2023evaluating, dai2023llmsinfoaccess},  the test set of Natural Questions (NQ) \cite{lee2019latent} dataset  which contains queries from Google search engine and answer from Wikipedia documents is selected in the section. 
GPT4o (temperature 0.5) is selected to rewrite the Wikipedia documents in different styles due to its good instruct-following ability and content generation ability \cite{islam2024gpt}. 
Given $N$ (query, gold document) pairs $[(q_1, d_1), (q_2, d_2), ..., (q_N, d_N) ]$ from NQ, the gold document is rewritten in nine writing styles by GPT4o, which transforms the original (query, gold document) pair $(q_i, d_i)$ into a set of 11 items including query, gold document and nine rewritten gold documents: 
$(q_i, d_i, d_i^{style0},  d_i^{style1}, ... , d_i^{style8})$.
The average document lengths are shown in Figure \ref{fig:stats} (a): Style 0 to 4 have similar text length as the original human written documents (dashed black line) while Style 5 to 8 have longer text lengths. Figure \ref{fig:stats} (b), (c), and (d) respectively illustrate the n-gram matching scores, as evaluated by Bleu, Meteor, and RougeL metrics, between the original human text and its nine stylistically rewritten counterparts. These scores share the similar conclusions: Styles 0 through 4, excluding Style 3, exhibit the greatest lexical similarity with the original text. This discrepancy with Style 3 is due to its informal nature, characterized by the use of emojis, abbreviations, and internet slang, which deviates significantly from the formal style of Wikipedia (original text). The n-gram matching scores decrease as one progresses from Style 5 to Style 8.

\begin{figure*}[h!]
  \centering
  \includegraphics[width=\linewidth]{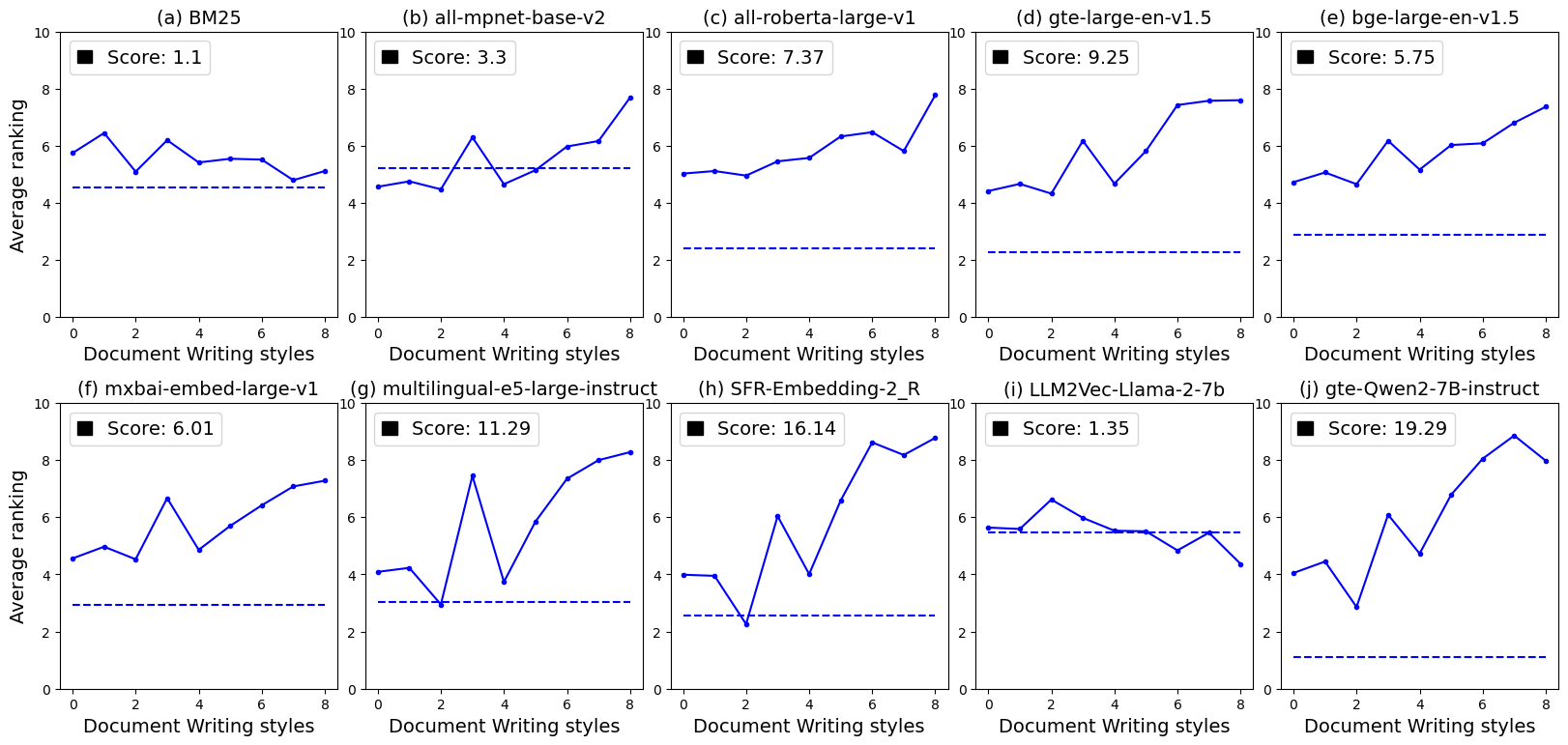}
  \caption{The original document as well as its 9 writing styles are ranked by their  similarity to the query  for each embedding model: the one with the highest similarity is assigned the lowest rank. In each subplot, the X-axis represents 9 writing styles (Style 0 to Style 8) while the Y-axis represents the value of average rank. Each writing style's average rank is denoted by a blue dot, whereas the average rank of the original human-written document  is indicated by a dashed line. The unfairness score (defined by Equation \ref{eq1}) of each model is also shown in each subplot. } 
  \label{fig:cc}
\end{figure*}

\textbf{Text embedding models:} Diverse top performing (on MTEB) universal text embedding models are investigated in this work, including:
\begin{itemize}
    \item BERT-based universal text embeddings: gte-large-en-v1.5 (434M parameters), bge-large-en-v1.5 (335M parameters), mxbai-embed-large-v1 (335M parameters), multilingual-e5-large-instruct (560M parameters).
    \item LLM-based universal text embeddings: SFR-Embedding-2\_R, LLM2Vec-Llama-2-7b, Qwen2-7B.
\end{itemize}
Three baseline models are selected including two widely used contextual text embedding models from SentenceTransformers\cite{sbertreimers2019sentence}: all-mpnet-base-v2 (110M parameters) and all-roberta-large-v1 (355M parameters), as well as one lexical retrieval model BM25 \cite{robertson2009probabilistic}.


\subsection{Experimental results on documents in different styles}

In order to study the bias and fairness of different text embedding models for information retrieval task, given ($q_i$, $d_i$, $d_i^{style-0}$,  $d_i^{style-1}$, $...$ , $d_i^{style-8}$), the cosine similarity is computed as in \cite{muennighoff2022mteb, dai2023llmsinfoaccess} between the text embeddings of the query $q_i$ and each of the ten documents $D_i$ = [$d_i$, $d_i^{style-0}$,  $d_i^{style-1}$, $...$ , $d_i^{style-8}$] which share the same semantic information but differ in writing styles.
Then, each document is ranked by its  similarity value to the query  for each embedding model: 
\begin{equation}\label{rank}
    R_{i} = Rank(\ [Cosine(Emb(q_i), \ Emb(d)), \ for\ d\ in\ D_i] \ )
\end{equation}
where $Emb()$ is a text embedding model, $Cosine( \cdot, \cdot)$ is Cosine similarity function. $Rank()$ is a ranking function: the document with the highest similarity to the query is assigned the lowest rank with value 1 while the one with the lowest similarity is assigned the largest rank value. Finally, the average rank of each writing style across the dataset is calculated:
\begin{equation}\label{avgrank}
    \overline{R} = \frac{1}{N} \times \Sigma_{i=1}^{N} R_{i} 
\end{equation}
where N is the datasize. $\overline{R}$ is a vector composed of the average ranking of the original document and nine different writing styles [$\overline{R}^d$,  $\overline{R}^{style0}$, $\overline{R}^{style1}$, ..., $\overline{R}^{style8}$].
If a text embedding model ensures fairness in information retrieval, the average ranking should be similar across various writing styles.

The average rankings of different writing styles across 10 different text embedding models are visualized in Figure \ref{fig:cc}. Each subplot represents the results of a text embedding model.  In each subplot, the X-axis represents 9 writing styles (Style 0 to Style 8) while the Y-axis represents the value of average rank. Each writing style's average rank is denoted by a blue dot, whereas the average rank of the original human-written document   is indicated by a dashed blue line.

\textbf{Is text generated by LLMs always preferred by embedding models?} Among 9 different writing styles, the Style-0 (default style) is from \cite{dai2023llmsinfoaccess}. The authors observed that lexical embedding model such as BM25 tend to favor human-written documents, while neural embedding models are biased towards LLM-generated Style-0 documents. In terms of BM25 model, similar behavior is observed in Figure \ref{fig:cc} (a): human-written text (dashed line in the figure) is ranked lower (favored) than all 9 different writing styles generated by GPT4o. However, Style-7 demonstrates an average ranking nearly identical to that of human-written documents.
Regarding neural embedding models, the state of the art universal text embeddings overlooked in \cite{dai2023llmsinfoaccess} are compared in this work. 
Style-0 is only favored by the baseline embedding model all-mpnet-base-v2 shown in Figure \ref{fig:cc} (b). Nevertheless, not every  LLM generated style is ranked lower (preferred) than human-written text by all-mpnet-base-v2, as seen in the case of Style-3, Style-6, Style-7, and Style-8.
For universal text embedding models: human written text is preferred (ranked lower) compared to Style-0.

\textbf{Which text embedding model is more biased?}
When the average rankings of different writing styles vary a lot for an embedding model, it indicates a bias towards a specific style in the model. In order to quantitatively measure the writing style bias, an unfairness score is proposed. Given a text embedding model's average rankings across  various writing styles $\overline{R}$ = [$\overline{R}^d$,  $\overline{R}^{style0}$, $\overline{R}^{style1}$, ..., $\overline{R}^{style8}$], the unfairness score is defined as:
\begin{equation}\label{eq1}
    Score = (max(\overline{R}) - min(\overline{R})) \times std(\overline{R})
\end{equation}
where $(max(\overline{R}) - min(\overline{R}))$ measures the spread or difference of rankings between the least preferred  and most preferred writing styles by the embedding model, $std(\overline{R})$ is the standard deviation which measures how much the rankings of different writing styles vary.
In essence, a larger score would indicate that a text embedding model is more biased and  has strong preference over certain writing style.
On the other hand, a smaller score would suggest that a text embedding model is more fair and have similar rankings across different writing styles. 

The unfairness scores of each model are also shown in Figure \ref{fig:cc}.
Most text embedding models are biased towards certain document writing style: all-roberta-large-v1, gte-large-en-v1.5,  bge-large-en-v1.5,
 mxbai-embed-large-v1, and gte-Qwen2-7B-instruct all exhibit the preference of human written text (dashed line in Figure \ref{fig:cc}).   
Among Style-0 to Style-4 which have similar text length as the original text,  most universal embedding  models show the least preference for  Style-3 (informal with emojis abbreviations, and internet slang).
Among Style-5 to Style-8 which have longer text length than the original document, Style-7 and Style-8 are least favored by these models.

The text embedding models with highest unfairness scores are gte-Qwen2-7B-instruct in Figure \ref{fig:cc} (j),  SFR-Embedding-2\_R in Figure \ref{fig:cc} (h) and 
multilingual-e5-large-instruct in Figure \ref{fig:cc} (g) even though these models have top performance across different tasks on MTEB.  
 gte-Qwen2-7B-instruct  demonstrates a strong bias towards original human-written text, as indicated by the near 1 average ranking of original documents (represented by the dashed line).
 SFR-Embedding-2\_R  and 
multilingual-e5-large-instruct  prefers most  Style-2, followed by the original documents:  among all text embedding models, only the fine-tuning data of  these two models contain the synthetic data generated by LLMs,  
this could explain why only these two models  show bias (preference) for Style-2.

\begin{figure}[h]
  \centering
  \includegraphics[width=\linewidth]{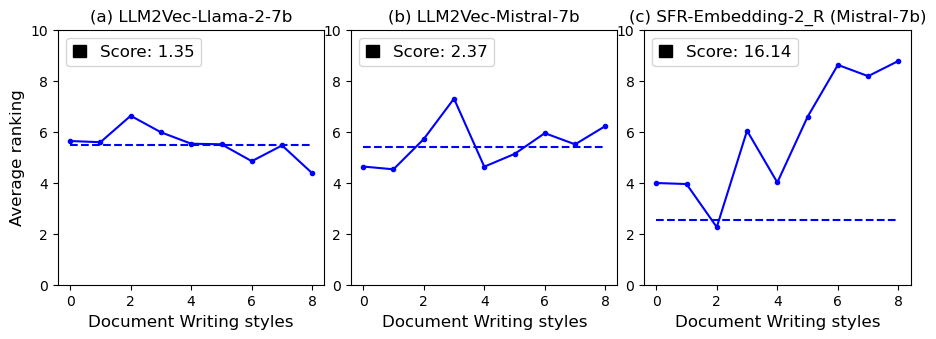}
  \caption{ The comparison of unfairness scores (defined by Equation \ref{eq1}) of LLM2Vec-Llama-2-7b, LLM2Vec-Mistral-7b and   SFR-Embedding-2\_R (Mistral-7b backbone). } 
  \label{fig:cc2}
\end{figure}

\textbf{Which text embedding model is more fair?}
The text embedding models with lowest unfairness scores are BM25 and LLM2Vec-Llama-2-7b. Both models have similar average rankings across all writing styles.
It is interesting to note that among all three LLM-based universal text embeddings, both SFR-Embedding-2\_R and gte-Qwen2-7B-instruct show strong bias towards certain writing styles while LLM2Vec-Llama-2-7b does not. These three embedding models differ in both the backbone LLM model and the fine-tuning method. In order to find out which factor plays a more important role in the model fairness, LLM2Vec-Mistral-7b based on Mistral-7b (same as SFR-Embedding-2\_R) is also tested. The comparison results are shown in Figure \ref{fig:cc2}: similar to LLM2Vec-Llama-2-7b, the unfairness score of LLM2Vec-Mistral-7b is also much lower than  SFR-Embedding-2\_R, even though they are based on the same backbone model. This finding suggests that the  fine-tuning strategy is more crucial for model fairness compared to the selection of backbone models for LLM-based universal text embeddings. 
The training recipe of LLM2Vec consists 4 steps: 1. enabling bidirectional attention for Decoder only LLMs, 2. enhancing bidirectional attention awareness with Masked Next Token Prediction (MNTP), 3. unsupervised contrastive learning, 4. supervised contrastive learning \cite{llm2vec}. Compared to other LLM-based embeddings, LLM2Vec devotes more efforts on enhancing bidirectional attention awareness of LLMs.

\subsection{Summary} 
In this section, ten different document writing styles (the original human written text and nine different generated writing styles) are compared  using diverse state of the art text embedding models. Through the experiments, it is observed that most universal embedding models exhibit bias towards certain document writing styles. The least favored writing styles include Style-3, Style-6, Style-7 and Style-8 which are more informal and emotive, while the most favored writing style including human written text (documents from Wikipedia) and Style-2 (clear and simple). Among all the MTEB top performing universal embedding models, the most biased are  gte-Qwen2-7B-instruct and SFR-Embedding-2\_R while the least biased is LLM2Vec-Llama-2-7b. The extended comparison with  LLM2Vec-Mistral-7b indicates that the fairness of LLM-based universal text embedding is more influenced by the fine-tuning strategy other than  the selection of the backbone LLM model.

\section{The bias and fairness related to query writing styles}

In the previous section, the bias and fairness of  text embedding models in terms of  document writing styles have been studied. However, variations in writing styles can also manifest in how individuals pose questions, influencing the outcomes of the IR system.  Therefore, the main focus of this section is  the bias and fairness of  text embedding models in terms of query writing styles.

\begin{figure}[t!]
  \centering
  \includegraphics[width=\linewidth]{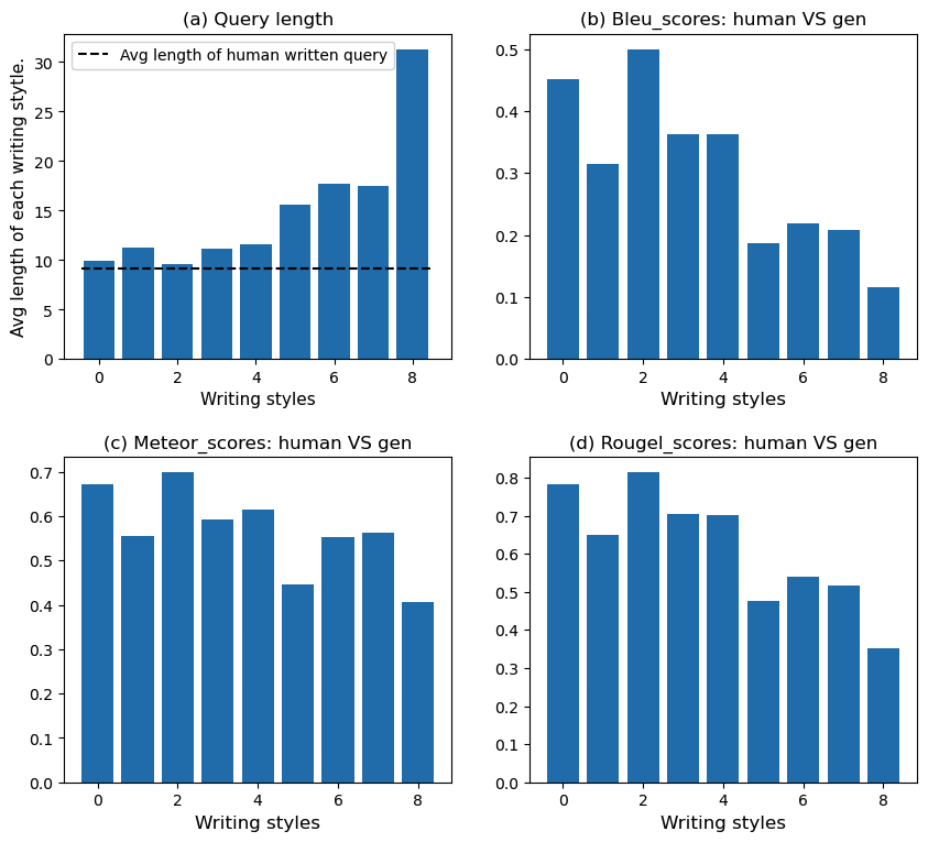}
  \caption{Basic statistics comparison between human written queries and generated queries in 9 different styles: (a) comparing the text length of different styles with the average length of the original query (dashed black line); (b),  (c)  and (d) are  Bleu scores, Meteor scores and RougeL scores  between original text and 9 writing styles.  } 
  \label{fig:stats2}
\end{figure}

The dataset remains unchanged from the previous section, with identical 9 writing styles employed on the query side in this section. Given the original query $q_i$, GPT4o is used to generate nine different writing styles: $[q_i^{style-0},  q_i^{style-1}, ... , q_i^{style-8}]$ which share the same semantic information as $q_i$. 
The average query lengths are shown in Figure \ref{fig:stats2} (a): Style 0 to 4 have similar text length as the original human written queries while Style 5 to 8 have longer text lengths. Figure \ref{fig:stats2} (b), (c), and (d) respectively illustrate the n-gram matching scores, as evaluated by Bleu, Meteor, and RougeL metrics, between the original human written queries and its nine stylistically rewritten counterparts. 

\begin{figure}[!]
  \centering
  \includegraphics[width=0.95\linewidth]{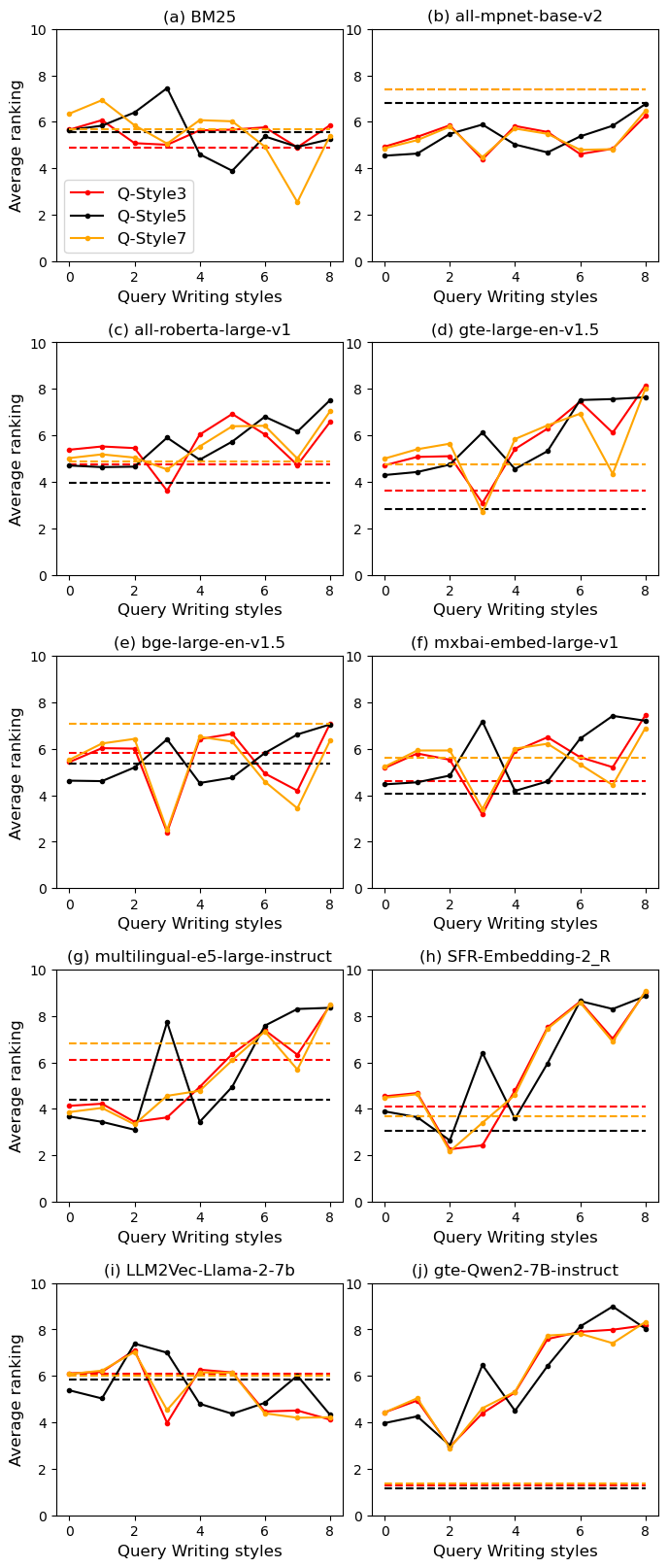}
  \caption{The impact of query writing styles on the IR fairness: X-axis represents 9 different query writing styles, Y-axis shows the average ranking. Query written in style 3, 5 and 7 are shown in color red, black and orange. Dashed line is the average ranking between query and the original human written document.} 
  \label{fig:qc1}
\end{figure}

In the previous section, given the original human written query $q$, the average rankings [$\overline{R}^d$,  $\overline{R}^{style0}$, $\overline{R}^{style1}$, ..., $\overline{R}^{style8}$] of the original human written document as well as its nine generated styles are calculated and compared.  In this section, in order to study the impact of query writing styles, the original human written query $q$ is replaced by its rewritten styles $q^{style-k}$ where $k \in [0,8]$. 
If a text embedding model ensures fairness and robustness in information retrieval, the average ranking should be similar across various document writing styles even though the query writing style changes.
The average rankings results for query style-3, query style-5 and query style-7 are shown with color red, black and orange respectively in  Figure \ref{fig:qc1} (results of rest 6 query styles  can be found in the Appendix \ref{app}). 

\begin{table*}[t!]
\caption{The unfairness scores for each embedding model on different query styles (Avg: average, Std: standard deviation).  }\label{t1}
\begin{adjustbox}{width=\textwidth}
\centering
\begin{tabular}{l p{1.2cm} p{1.2cm} p{1.2cm} p{1.2cm} p{1.2cm} p{1.2cm} p{1.2cm} p{1.2cm} p{1.2cm} p{1.2cm} p{.6cm} p{.6cm}  }
\toprule
models & Query (original) & Query (style-0) &  Query (style-1) &  Query (style-2) &  Query (style-3) &  Query (style-4) &  Query (style-5) &  Query (style-6) &  Query (style-7) &  Query (style-8) & Avg & Std\\ 
\midrule
BM25 & 1.10 & 1.12 & 2.16 & 1.00 & 0.53 & 1.76 & 3.48 & 2.92 & 5.23 & 9.30 & 2.85 & 2.53 \\
all-mpnet-base-v2 & 3.30 & 1.72 & 1.95 & 1.70 & 2.71 & 1.85 & 1.88 & 4.05 & 2.60 & 7.32 & 2.98 & 1.64 \\
all-roberta-large-v1 & 7.37 & 4.11 & 4.64 & 3.96 & 3.20 & 4.40 & 3.97 & 0.73 & 2.06 & 1.14 & 4.49 & 1.85 \\
gte-large-en-v1.5 & 9.25 & 9.16 & 9.27 & 9.51 & 7.95 & 9.02 & 7.93 & 2.47 & 7.81 & 2.16 & 8.29 & 2.65 \\
bge-large-en-v1.5 & 5.75 & 2.87 & 3.10 & 3.08 & 6.39 & 2.80 & 2.34 & 3.18 & 6.94 & 6.16 & 4.66 & 1.71 \\
mxbai-embed-large-v1 & 6.01 & 4.85 & 5.16 & 4.96 & 4.79 & 4.99 & 4.60 & 1.96 & 3.42 & 5.50 & 4.99 & 1.09 \\
multilingual-e5-large-inst & 11.29 & 13.77 & 13.69 & 14.46 & 8.44 & 13.58 & 11.65 & 2.68 & 8.72 & 4.05 & 11.15 & 3.98 \\
SFR-Embedding-2\_R & 16.14 & 16.56 & 16.10 & 17.04 & 16.34 & 16.20 & 15.19 & 9.67 & 16.14 & 7.21 & 16.59 & 3.24 \\
LLM2Vec-Llama-2-7b & 1.35 & 1.70 & 2.30 & 1.20 & 3.46 & 2.10 & 3.22 & 6.94 & 2.94 & 11.79 & 3.67 & 3.12 \\
gte-Qwen2-7B-instruct & 19.29 & 19.53 & 19.83 & 19.34 & 16.28 & 19.68 & 19.85 & 15.32 & 16.03 & 16.04 & 21.16 & 2.04 \\
\bottomrule
\end{tabular}
\end{adjustbox}
\end{table*}

\textbf{Do embedding models match document style with the query style?} 
In the previous section, given the original human written query, most text embedding models prefer more the original human written documents while prefer less style-3 (informal with emojis) documents (among style-0 to style-4).    When the query is also written in style-3, the results are shown in color red in Figure  \ref{fig:qc1}: most embedding models match the document writing style with the query writing style and prefer most style-3 document apart from multilingual-e5-large-instruct (Figure \ref{fig:qc1} (g)), SFR-Embedding-2\_R (Figure \ref{fig:qc1} (h)), and gte-Qwen2-7B-instruct (Figure \ref{fig:qc1} (j)). Similar observations can be found when changing the query style to  style-5 (black lines in 
Figure \ref{fig:qc1}) and style-7 (orange lines in Figure \ref{fig:qc1}). These observations suggest that most text embedding models prefer the document style that is the same to the query style. 

 One exception is gte-Qwen2-7B-instruct (Figure \ref{fig:qc1} (j)): no matter how the query style changes, gte-Qwen2-7B-instruct always demonstrates a strong bias towards original human-written text (prefers most  the original human written documents as in the previous section).    
For two embedding models find-tuned with GPT3.5/GPT4 data, multilingual-e5-large-instruct (Figure \ref{fig:qc1} (g)) and SFR-Embedding-2\_R (Figure \ref{fig:qc1} (h)) always  prefer the Style-2  documents regardless of changes in the query style.

\textbf{Which text embedding model is more fair and robust across different query styles?} The unfairness score defined in  Equation \ref{eq1} is calculated for each query style in this section (shown in Table \ref{t1}). Even though universal text embeddings have better performances than three baseline methods (BM25, all-mpnet-base-v2 and all-roberta-large-v1) across various tasks on MTEB benchmark, these baselines have smaller average unfairness scores compared to universal embeddings.  
Similar to the findings from the previous section,  among the state of the art universal text embeddings, LLM2Vec-Llama-2-7b  exhibit more fairness across different writing styles. 
However, mxbai-embed-large-v1 is the most robust model over different query styles (with the smallest Std in Table \ref{t1}). That is because BM25, all-mpnet-base-v2 and LLM2Vec-Llama-2-7b have larger unfairness score on Query style-8 even though their average unfairness scores are low. 
The most biased embedding models are  gte-Qwen2-7B-instruct, SFR-Embedding-2\_R and multilingual-e5-large-instruct. gte-Qwen2-7B-instruct
demonstrates a strong bias towards original human-written text, while SFR-Embedding-2\_R and multilingual-e5-large-instruct shows a pronounced preference for Style-2 regardless of the query style change.

\textbf{Summary:} In this section, the impact of query writing styles on the fairness of IR systems is discussed. Most embedding models  match the document writing style with the query writing style. This means that when query is asked in a certain style, the  response retrieved by most text embedding models is highly probably to be in the same style. However, there are several exceptions:  multilingual-e5-large-instruct  and SFR-Embedding-2\_R always prefer the Style-2 documents regardless of changes in the query style, while gte-Qwen2-7B-instruct has a strong preference over the original human-written text. Among all the universal text embedding models, the most fair and robust ones are LLM2Vec-Llama-2-7b and mxbai-embed-large-v1.

\section{The impact of LLM's answer styles}
The preceding sections explored how document and query writing styles (from diverse human profiles) influence IR system fairness, in order to study if certain user groups are discriminated by IR systems based on the state of the art text embedding models. Recently,  LLM-based RAGs have gained popularity in Question-Answering (QA) across different domains \cite{chen2024spiral}. In this section, we study whether RAGs based on different LLMs have different answer styles (under the same prompt) and what are  the potential impacts of these answer styles when text embedding similarity is used as evaluation metrics of answer correctness.

\subsection{Data}
The Instruct-QA dataset \cite{adlakha2024evaluating} is used in this section: Instruct-QA dataset contains on three diverse information-seeking QA tasks including Natural Questions \cite{lee2019latent}, HotpotQA \cite{yang2018hotpotqa} and TopiOCQA \cite{adlakha2022topiocqa}. RAGs based on 4 different LLMs including FlanT5-xxl \cite{chung2024scalingt5}, Alpaca-7b \cite{taori2023stanfordalpaca}, GPT3.5-turbo and Llama2-7b \cite{touvron2023llama} are used to provide answers and each answer's correctness is annotated by human evaluators. 
Only LLM's answers that are annotated as correct are selected. 

\begin{figure}[t!]
  \centering
  \includegraphics[width=0.9\linewidth]{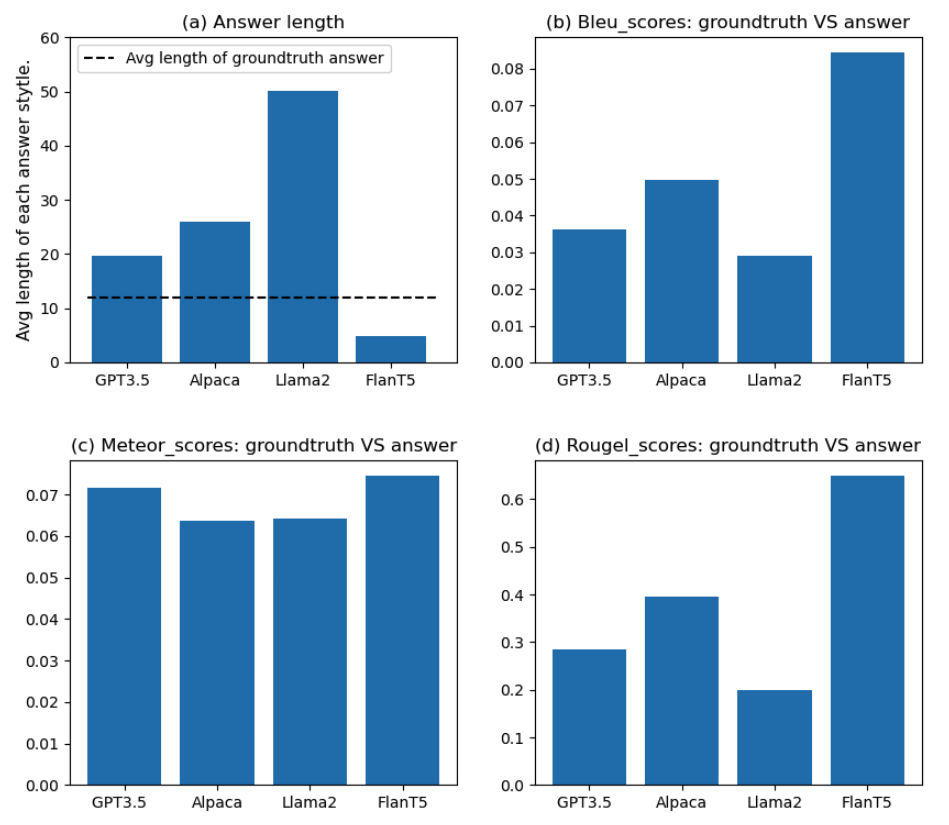}
  \caption{Basic statistics of answers generated by different LLMs: (a) comparing the answer length of different LLMs with the average length of the Ground-Truth (GT) answer (dashed black line);  (b),  (c)  and (d) are  Bleu scores, Meteor scores and RougeL scores  between GT and  answers from different LLMs.  } 
  \label{fig:stats3}
\end{figure}

The answer lengths from different LLMs are shown in Figure \ref{fig:stats3} (a) and the Ground-Truth (GT) answer length is the dashed black line: Llama2-7b has the most verbose answer, followed by Alpaca-7b and GPT3.5-turbo, while FlanT5-xxl provides the most brief answer.  
The n-gram matching scores including Bleu, Meteor and RougeL scores between GT and each answer are shown in \ref{fig:stats3} (b), (c), (d) respectively. For  Meteor scores, the  answers from 4 LLMs have similar lexical similarity with the GT. For Bleu and RougeL scores, the answers provided by FlanT5-xxl exhibit the greatest degree of lexical similarity with the GT, followed by Alpaca-7b. The  Llama2-7b  answer have the least lexical similarity with the GT. 
\textbf{These findings  illustrate that the responses given by RAGs based on various LLMs exhibit a wide range of styles too.}
Here is one example of answers from RAGs based on different LLMs:
\begin{itemize}
    \item Question: What Theo Avgerinos movie did the actor known for the role as Matt McNamara appear in?
    \item Ground-Truth answer:  Fifty Pills
    \item GPT3.5 answer:   John Hensley appeared in Theo Avgerinos' movie "Fifty Pills".
    \item Alpaca answer:   The actor known for the role as Matt McNamara appeared in the movie "Fifty Pills".
    \item Llama2 answer: The actor known for his role as Matt McNamara in "Nip/Tuck" appeared in "Fifty Pills" directed by Theo Avgerinos.
      \item FlanT5 answer:   Fifty Pills
\end{itemize}

\subsection{Experimental results}

\begin{figure}[!]
  \centering
  \includegraphics[width=0.9\linewidth]{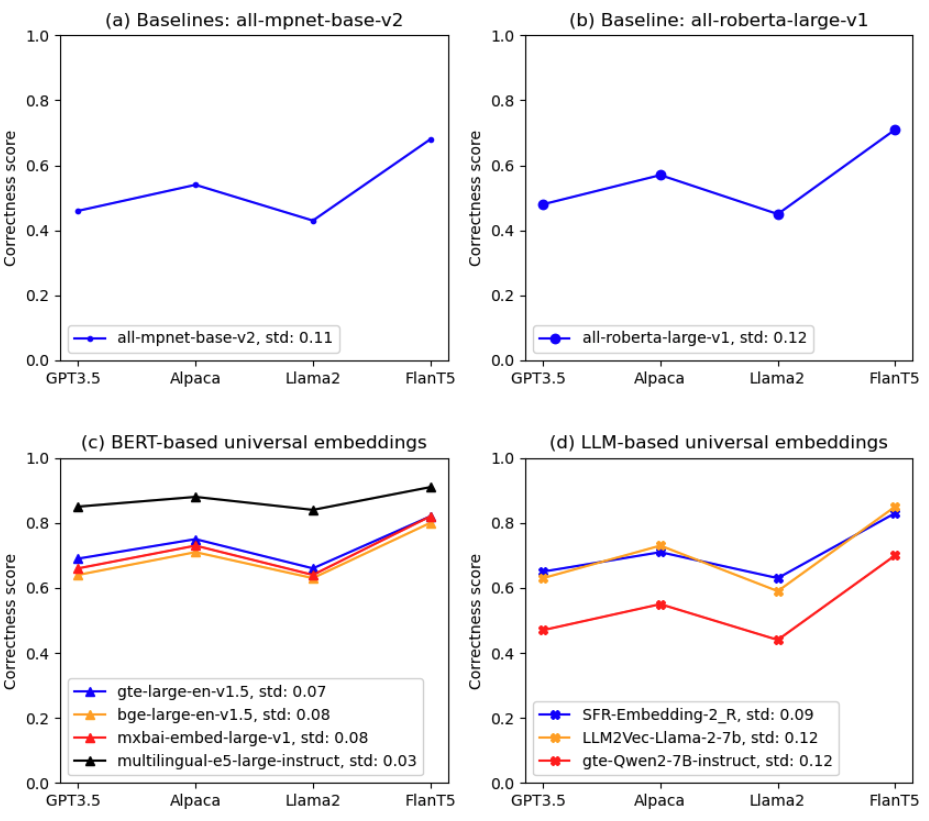}
  \caption{The impact of LLM's answer writing styles on the correctness evaluation of IR systems based on text embeddings: X-axis represents answers from  four different LLMs, Y-axis shows the correctness score. The unfairness score of each embedding model is also presented in each subplot.} 
  \label{fig:llm}
\end{figure}

In IR systems, text embeddings are also used as an evaluation metric to measure the answer correctness \cite{es2023ragas}. 
The text embedding similarity between the GT and RAG's answer is frequently used as the correctness score. 
In order to study the impact of answer styles on the correctness evaluation, we calculate correctness score of each LLM using the semantic similarity between GT answer and  LLM's answer based on different text embeddings. 
Since all the LLMs' answers are annotated as correct in this experiment, differing only in answer styles, their correctness score should be similar. 

 The  correctness scores based on different text embedding models are shown in Figure \ref{fig:llm} (BM25 is excluded as its values are unbounded): multilingual-e5-large-instruct shown in black color in Figure \ref{fig:llm} (c) demonstrates the most fairness across different answer styles (with 0.03 std). However, most models show a tendency to assign higher correctness scores to FlanT5-xxl's answers, while giving lower scores to Llama2-7B and GPT3.5-turbo. This is similar to the patterns observed in the Bleu and RougeL scores shown in Figure \ref{fig:stats3}. These results indicate a prevalent bias in text embedding models towards LLM's answer styles when used as evaluation metrics for answer correctness.

\section{Conclusion}
In this work, we initiate the investigation of the potential biases in state-of-the-art text embedding models towards certain document and query writing styles. 
We observe that most top-performing universal embedding models tend to favor certain document writing style, with a particular preference for clear, simple styles and human-written Wikipedia style text. Conversely, more informal and emotive styles are least favored. 
In terms of query writing styles, the majority of embedding models tend to match the style of the query with the style of the retrieved documents. However, some models exhibit a consistent preference for specific styles, regardless of the query style. 
Text embedding models fine-tuned on synthetic data generated by LLMs are found to have a consistent preference over certain style of generated data (Style-2). 
These biases in text embedding based IR systems can inadvertently silence or marginalize certain communication styles, thereby posing a significant threat to fairness in information retrieval.

In light of these findings, future work should prioritize the development of fairer, more balanced text embedding models that do not favor or marginalize certain writing styles. 
Furthermore, we advocate for greater transparency and awareness about these biases among users and developers of information retrieval systems. This could involve providing clear information about the potential biases of different models, and offering users the option to choose the model that best suits their needs and preferences.

\section{Ethical Considerations}
This study investigates the potential bias and fairness of Information Retrieval Systems, utilizing the state of the art text embedding models, without the necessity of releasing the generated text or the interference of societal advancement. To carry out our research, we utilized Language Learning Models (LLMs) and data sets that are publicly accessible, ensuring our experiments did not raise any ethical concerns.

\bibliographystyle{ACM-Reference-Format}
\balance
\bibliography{sample-base}

\appendix

\section{APPENDIX} \label{app}
\begin{figure}[H]
  \centering
  \includegraphics[width=0.9\linewidth]{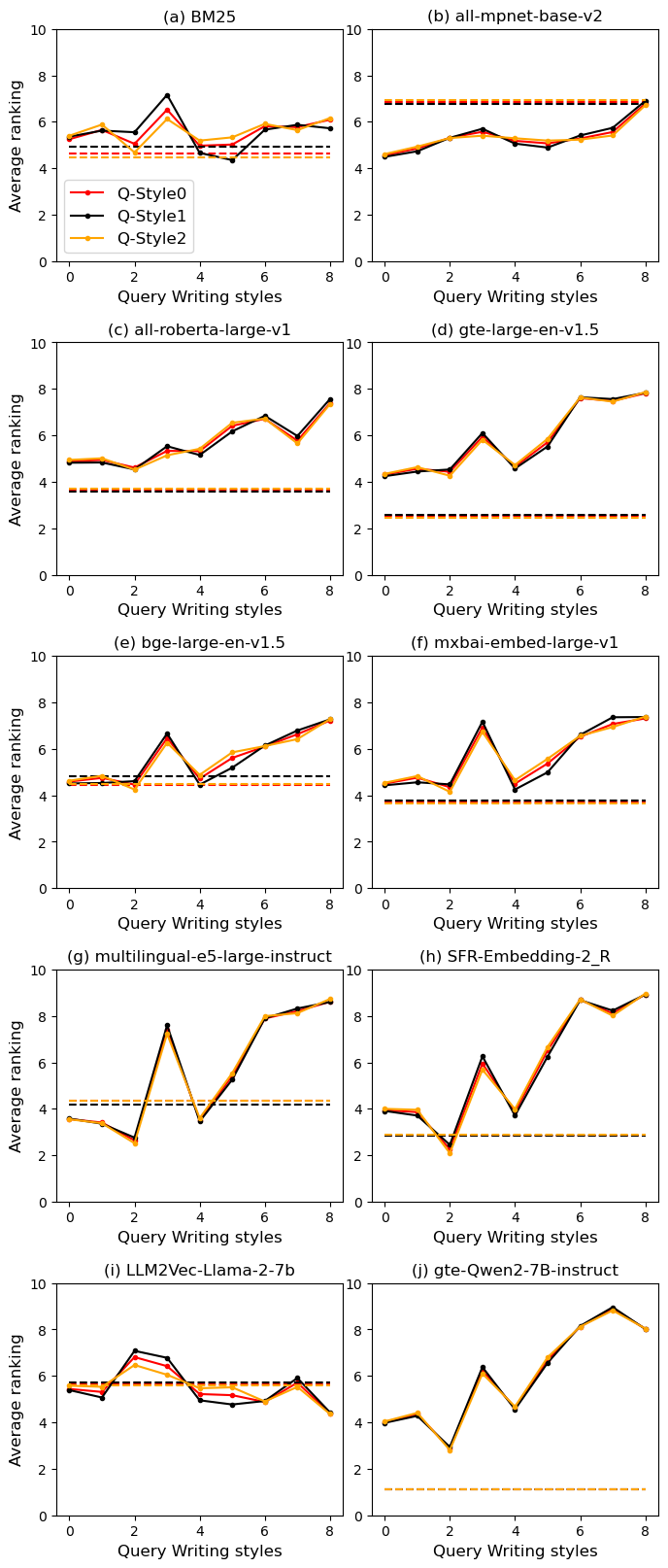}
  \caption{The impact of query writing styles on the IR fairness: X-axis represents 9 different query writing styles, Y-axis shows the average ranking. Query written in style 0, 1 and 2 are shown in color red, black and orange. Dashed line is the average ranking between query and the original human written document.} 
  \label{fig:qc0}
\end{figure}

\begin{figure}[H]
  \centering
  \includegraphics[width=0.9\linewidth]{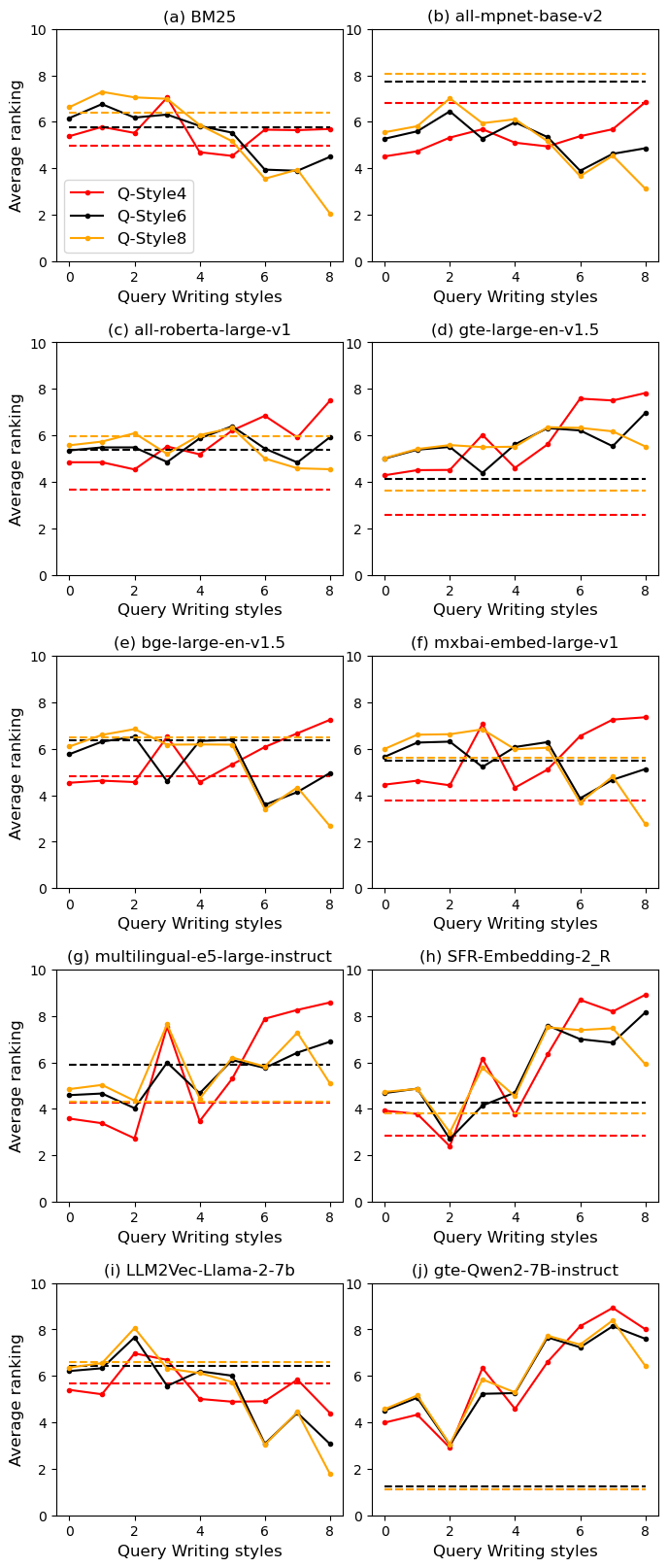}
  \caption{The impact of query writing styles on the IR fairness: X-axis represents 9 different query writing styles, Y-axis shows the average ranking. Query written in style 4, 6 and 8 are shown in color red, black and orange. Dashed line is the average ranking between query and the original human written document.} 
  \label{fig:qc2}
\end{figure}

\end{document}